\documentclass[aps,tightenlines,showpacs,nofootinbib,superscriptaddress]{revtex4}
\usepackage{graphicx}
\usepackage{amsmath}
\usepackage{amsfonts}
\usepackage{amssymb}
\usepackage{bm}

\begin{document}
\title{Ordering in Two-Dimensional Ising Models with
Competing Interactions}
\author{Gennady Y. Chitov}
\affiliation{Department 7.1-Theoretical Physics, University of
Saarland, Saarbr\"ucken D-66041, Germany} \affiliation{Department of
Physics and Astronomy, Laurentian University, Sudbury, ON, P3E 2C6
Canada}
\author{Claudius Gros}
\affiliation{Department 7.1-Theoretical Physics, University of
Saarland, Saarbr\"ucken D-66041, Germany}

\date{\today}

\begin{abstract}
We study the 2D Ising model on a square lattice with additional
non-equal diagonal next-nearest neighbor interactions. The cases of
classical and quantum (transverse) models are considered. Possible
phases and their locations in the space of three Ising couplings are
analyzed. In particular, incommensurate phases occurring only at
non-equal diagonal couplings, are predicted. We also analyze a
spin-pseudospin model comprised of the quantum Ising model coupled
to $XY$ spin chains in a particular region of interactions,
corresponding to the Ising sector's super-antiferromagnetic (SAF)
ground state. The spin-SAF transition in the coupled Ising-XY model
into a phase with co-existent SAF Ising (pseudospin) long-range
order and a spin gap is considered. Along with destruction of the
quantum critical point of the Ising sector, the phase digram of the
Ising-XY model can also demonstrate a re-entrance of the spin-SAF
phase. A detailed study of the latter is presented. The mechanism of
the re-entrance, due to interplay of interactions in the coupled
model, and the conditions of its appearance are established.
Applications of the spin-SAF theory for the transition in the
quarter-filled ladder compound $\rm NaV_2O_5$ are discussed.
\end{abstract}
\pacs{ 71.10.Fd, 71.10.Hf, 75.30.Et, 64.60.-i}

\maketitle
%
%
\section{Introduction}\label{Intro}
%
%
The role of competing interactions in ordering is a fascinating
problem of condensed matter physics. One of the most canonical
examples of such systems are frustrated Ising models which
demonstrate a plethora of critical properties, far from being
exhaustively studied. (For a review see \cite{Liebmann86}.) The
frustrations\footnote{ \label{noteFr} We use the term ``frustration"
in a broad sense \cite{Liebmann86}, meaning only that there is no
spin arrangement on an elementary plaquette which can satisfy all
bonds.} can be either geometrical, like, e.g., in the Ising model on
a triangular lattice, or they can can be brought about by the
next-nearest neighbor (nnn) interactions. Competing interactions
(frustrations) can, e.g., result in new phases, change the Ising
universality class, or even destroy the order at all. Another
interesting aspect of the criticality in frustrated Ising models is
an appearance of Quantum Critical Point(s) (QCP) at special
frustration points of model's high degeneracy, and related quantum
phase transitions \cite{SachdevQPT}.

Inclusion of a transverse field ($\Omega$) brings an extra scale
into the game, giving raise to a new and complicated critical
behavior. The Ising models with $\Omega=0$ and $\Omega \neq 0$ are
also often called classical and quantum, respectively. For a review
on the Ising Models in Transverse Field (IMTF) see \cite{Chak96}.
Most studies of the frustrated quantum Ising models are restricted
to their ground states properties, when mapping of the
$d$-dimensional quantum model at $T=0$ onto its ($d+1$)-dimensional
classical counterpart helps to analyze the ground state phase
diagram of the former.  For the nnn 2D models we are interested in,
there has been a considerable effort on the quantum ANNNI model,
reviewed in \cite{Chak96}. The studies of some other 2D frustrated
transverse Ising models have appeared only recently
\cite{Moessner01,Most03}.

Our interest in the subject comes from the earlier work on a quantum
Ising model coupled to the spin chains \cite{CGbig03}. This kind of
coupled so-called spin-pseudospin (or spin-orbital) models appear in
context of the phase transition in $\rm NaV_2O_5$, which has
inspired a great experimental and theoretical effort in recent
years. (For a review see \cite{Lem03}.) Going deeper into analysis,
we came to realize that the Ising sector of the problem is in fact
the 2D transverse nn and nnn Ising model on a square lattice. It
turns out that even its classical counterpart ($\Omega=0$) was
studied only for the case of equal nnn couplings $J_1=J_2$
\cite{Liebmann86}. The elementary plaquette of this lattice with the
notations for couplings is shown in Fig.\ \ref{IsNNN}. To the best
of our knowledge, this 2D nn and nnn Ising model in transverse field
is a complete \textit{terra incognita} even at $J_1=J_2$.

So, as the first step, we find the ground state phase diagram of the
classical ($\Omega=0$) Ising model at arbitrary couplings
$J_{\text{\tiny $\square$}}, J_1,J_2$. Along with the three ordered
phases found earlier by Fan and Wu \cite{FanWu69} for the case
$J_1=J_2$ --ferromagnetic (FM), antiferromagnetic (AF), and
super-antiferromagnetic (SAF)-- the model has a fourth phase which
can occur if $\textrm{sign}(J_1/J_2)=-1$. We call it
super-ferro-antiferromagnetic (SFAF) \cite{Landau85}. From
mean-field-type arguments we predict also the existence of an
incommensurate (IC) phase at $T >0$ in this model. The 2D IC phase
is also called floating \cite{Bak82}. Similar phase is known for the
well-studied 2D ANNNI model \cite{Liebmann86,Bak82,Selke88}. We
present a qualitative temperature phase diagram for the regions of
the coupling space where the IC phase is located. Note that the
three phases -- SAF, SFAF, IC -- can occur only in the presence of
competing interactions in the Ising model, and the latter two occur
only if $J_1 \neq J_2$.

This analysis of the classical Ising model lays the grounds for
venturing into its study in presence of a transverse field. The role
of transverse field is subtle. A more straightforward aspect is that
its increase above certain critical value can eventually destroy the
ordered state of the classical model, and in the ground state the
transverse field results in appearance of a QCP. This is similar to
the well-understood quantum nn Ising model. Another particularly
interesting aspect in the role of transverse field is that it can
lift degeneracy of the ground state and stabilize new phases at
finite temperature in a highly frustrated model, like, e.g., the
antiferromagnetic isotropic triangular Ising model
\cite{Moessner01,Most03}, which is disordered at any $T>0$ when
$\Omega=0$.

The behavior of the systems with infinitely degenerate ground states
(with or without a finite ground-state entropy per spin) can be quite
complicated in the presence of transverse field. It lies
beyond the scope of the present work, and definitely cannot be understood
from the mean-field analysis we apply in this study. For the nn and nnn
Ising model we only identify the lines (planes) in the space of couplings
($J_{\text{\tiny $\square$}}, J_1,J_2$) where the model is highly degenerate,
and in their neighborhood we expect some new exotic phases generated by
$\Omega \neq 0$ to appear.

From mapping of the nn and nnn IMTF at $T=0$ onto its classical 3D
counterpart, we qualitatively predict the (mean-field) ground-state
phase boundaries of the quantum model in the coupling space
$(J_{\text{\tiny $\square$}}, J_1,J_2)$. In particular, it follows
from our analysis that in the presence of transverse field  the IC
ground-state phase can penetrate into some parts of the FM, AF, SAF
regions of the classical model ($\Omega=0$).

Finally, we consider the coupled spin-pseudospin model. It is
proposed to analyze the transition in $\rm NaV_2O_5$. This material
provides a unique example of a correlated electron system, where the
interplay of charge and spin degrees of freedom results in a phase
transition into a phase with coexistent spin gap and charge order.
$\rm NaV_2O_5$ is the only known so far quarter-filled ladder
compound. Each individual rung of a ladder is occupied by single
electron which is equally distributed between its left/right sites
in the disordered phase. At $T_c=34\,{\rm K}$ this compound
undergoes a phase transition when a spin gap opens, accompanied by
charge ordering \cite{Lem03}.

The problem of the electrons in $\rm NaV_2O_5$ localized on the
rungs of the 2D array of ladders is mapped onto the coupled
spin-pseudospin model on the effective square lattice. The Ising
sector of this model is given by the Hamiltonian of the nn and nnn
IMTF, and the Ising variables (called pseudospins for this case)
represent physically the charge degrees of freedom. We model the
spin sector by the array of the $XY$ spin chains. The 2D long-range
charge order in $\rm NaV_2O_5$ is identified as the SAF phase of the
Ising model,  and we restrict our analysis to the SAF region of the
couplings $(J_{\text{\tiny $\square$}}, J_1,J_2)$. The coupled model
is handled by combining the mean-field treatment of its Ising sector
with the use of exact results available for the $XY$ spin chains.
Since the SAF state is only four-fold degenerate, the mean-field
predictions for the Ising sector of the coupled model are expected
to be at least qualitatively correct.

The mean-field equations for the coupled model are, with some minor
modifications, the same as we have obtained earlier \cite{CGbig03}.
A striking feature of the coupled model is that it \textit{always}
orders from the charge-disordered spin-gapless state into the phase
of co-existent SAF charge order and spin gap. We call this the
spin-SAF transition. By \textit{always} we mean that the critical
temperature of the spin-SAF transition is non-zero for all Ising
couplings within the whole considered SAF region. In other words,
the QCP of the IMTF is destroyed, and this is due to the spin-charge
(-pseudospin) coupling. This property of the spin-SAF transition and
the parameters of the spin-SAF phase were studied earlier
\cite{CGbig03}, so in this work we only reinstate some points and
stress the distinctions pertinent to the present model.

The other remarkable feature of the coupled model's phase diagram is
re-entrance, which was not well understood in our earlier work
\cite{CGbig03}. Now we carry out an analytical study of the
re-entrance and establish the conditions when it can occur. This
analysis allows us to understand the detailed mechanism of this
interesting phenomenon generated by competing interactions.

The rest of the paper is organized as follows. Section \ref{NNNI}
contains our results on the ordering in the 2D nn and nnn Ising
model at $\Omega=0$ and $\Omega \neq 0$. The results on the spin-SAF
transition in the coupled model are presented in Section \ref{SPM}.
The final Section \ref{Concl} presents the summary and discussion.
%
%
%
\section{2D Nearest- and next-nearest-neighbor Ising Model}\label{NNNI}
We consider the 2D Ising Model on a square lattice with the
Hamiltonian
\begin{equation}
\label{NNNIs}
H =
\frac12 \sum_{\langle \mathbf{i},\mathbf{j} \rangle}
J_{\text{\tiny $\square$}}
\mathcal T^x_{\mathbf{i}} \mathcal T^x_{\mathbf{j}}
+\frac12 \sum_{\langle \langle \mathbf{k},\mathbf{l} \rangle \rangle}
J_{\mathbf{k} \mathbf{l}}
\mathcal T^x_{\mathbf{k}} \mathcal T^x_{\mathbf{l}}
\end{equation}
where the bold variables denote lattice vectors, the first [second]
sum includes only nearest neighbors (nn) [next-nearest neighbors
(nnn)] of the lattice, respectively. Spins along the sides of an
elementary plaquette interact via the nn coupling $J_{\text{\tiny
$\square$}}$, while spins along plaquette's diagonals interact via
the nnn couplings $J_{\mathbf{k} \mathbf{l}}=J_{1,2}$ (see Fig.\
\ref{IsNNN}). The way we defined the Hamiltonian corresponds to
antiferromagnetic couplings for $J_\sharp
>0$ and ferromagnetic for $J_\sharp <0$.
%
\subsection{Ground state phases}\label{GSph}
\begin{figure*}
\includegraphics[width=8.5cm]{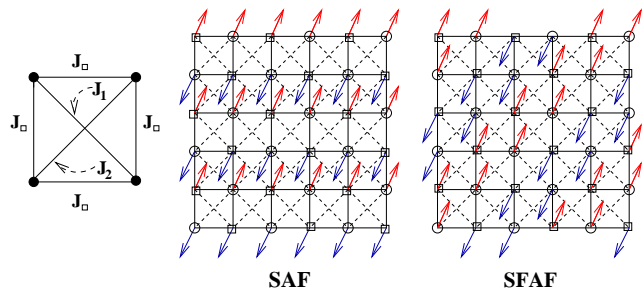}
\caption{Left: Couplings on an elementary plaquette in the nn and
nnn 2D Ising model (\ref{NNNIs}). Ordering patterns in the
super-antiferromagnetic (SAF) and super-ferro-antiferromagnetic
(SFAF) phases. } \label{IsNNN}
\end{figure*}

%
There is no exact solution of the model (\ref{NNNIs}). Its possible
ordered phases and critical properties have been studied within
various approaches for equal diagonal couplings $J_1=J_2$ (see
\cite{Liebmann86} for a review and references on the original
literature). We will consider arbitrary Ising couplings
$(J_1,J_2,J_{\text{\tiny $\square$}})$, so the model (\ref{NNNIs})
can be either frustrated or not (see footnote \ref{noteFr}). The
ground state phase diagram can be found from energy arguments, as
was first done by Fan and Wu for $J_1=J_2$ \cite{FanWu69}. (Their
phase diagram is shown in Fig.\ \ref{PDcut}c.) From direct counting
of the ground state energies of possible spin arrangements we
construct the phase diagram for $J_1 \neq J_2 $. Along with the
phases found by Fan and Wu-- ferromagnetic (FM), antiferromagnetic
(AF), and super-antiferromagnetic (SAF)-- there is a fourth phase
which can occur if $J_1$ and $J_2$ have opposite signs. The name of
the SAF phase comes from viewing it as two superimposed
antiferromagnetic lattices (one lattice of circled sites and another
of squared sites in Fig.\ \ref{IsNNN}). In the SAF state there are
two frustrated bonds $J_{\text{\tiny $\square$}}$ per plaquette and
its energy is four-fold degenerate, since each of the superimposed
lattices can be flipped independently. In addition to the two SAF
states with alternating ferromagnetic order along the horizontal
chains (one of these is shown in Fig.\ \ref{IsNNN}), there are two
states with the vertical ferromagnetic order.

The new fourth phase shown in Fig.\ \ref{IsNNN} can be viewed as two
superimposed lattices each of which is ordered ferromagnetically
along one side (e.g., $J_2<0$) and antiferromagnetically along the
other (e.g., $J_1>0$). So we will call it
super-ferro-antiferromagnetic (SFAF) \cite{Landau85}. The SFAF state
also has two frustrated plaquette's bonds and four-fold degeneracy.
The ordering pattern shown in Fig.\ \ref{IsNNN} changes only by a
lattice spacing shift over flipping of the sublattices. The
direction of the ferromagnetic order is determined by the
ferromagnetic diagonal.

The ground state phases in the space $(J_1,J_2,J_{\text{\tiny
$\square$}})$ are shown in Fig.\ \ref{PD3DFul}. In order to
facilitate perception of this picture, we also present in Figs.\
\ref{PDcut},\ref{PDpl} several plane projections of the 3D Fig.\
\ref{PD3DFul}. In the first quadrant ($J_1,J_2 > 0$) in the region
$J_1+J_2 > |J_{\text{\tiny $\square$}}|$ lying between two
frustration planes FP
\begin{equation}
\label{FP}
J_1+J_2 = |J_{\text{\tiny $\square$}}|: \quad \textrm{FP}
\end{equation}
the ground state of the model is SAF. A continuous transition from
the paramagnetic (PM) to the SAF phase occurs at some critical
temperature $T_c>0$. From the arguments known for the case $J_1=J_2$
\cite{KriMu77} (see also \cite{Dom78} for a more general symmetry
analysis of the Ginzburg-Landau functional) this transition is
non-universal: the critical indices continuously depend on the
couplings $J_{\text{\tiny $\square$}}, J_1,J_2$.

In the region $J_1>J_{\text{\tiny $\square$}}$ of the fourth quadrant
($J_1>0,J_2 < 0$) lying between the other pair of frustration planes FP'
\begin{equation}
\label{FP'}
J_1 = |J_{\text{\tiny $\square$}}|: \quad \textrm{FP'}
\end{equation}
the ground state is SFAF. The same arguments \cite{KriMu77,Dom78}
suggest non-universality of the PM $\to$ SFAF transition.

In the regions lying above (beneath) the frustration planes FP and
FP', and above (beneath) the basal plane $J_{\text{\tiny
$\square$}}=0$ in the third quadrant, the ground state is a usual AF
(FM), respectively. The transition PM $\to$ AF (FM) belongs to the
2D Ising universality class. The second quadrant $J_1<0,J_2
> 0$, not shown in Fig.\ \ref{PD3DFul}, is obtained by a reflection
over the plane $J_1=J_2$. In the AF (FM) state the number of
frustrated (diagonal) bonds per plaquette is two in the first
quadrant, one in the second and the fourth, and zero in the third.

Transitions at finite temperature should be absent on the
frustration planes FP/FP' where the model is highly degenerate. We
are not aware of studies of the ground state in these cases and
cannot say at the moment whether the system possesses some kind of a
long-range order at zero temperature or not, except a rather trivial
line $J_{\text{\tiny $\square$}}=J_1=0,J_2<0$ of the FP' planes
crossing where the model becomes a set of decoupled Ising chains,
and four special lines on the FP planes where it becomes the
exactly-solvable Isotropic Triangular Ising (ITI) model. The latter
case will be discussed momentarily.
%
\subsection{Exactly-solvable limits}\label{ExLim}
%
In the 3D space ($J_1,J_2,J_{\text{\tiny $\square$}}$) beside the
frustration planes FP and FP', there are three special planes where
one of the couplings is zero. On these planes the model
(\ref{NNNIs}) reduces to the exactly solvable cases.

\begin{figure*}
\includegraphics[width=5.5cm]{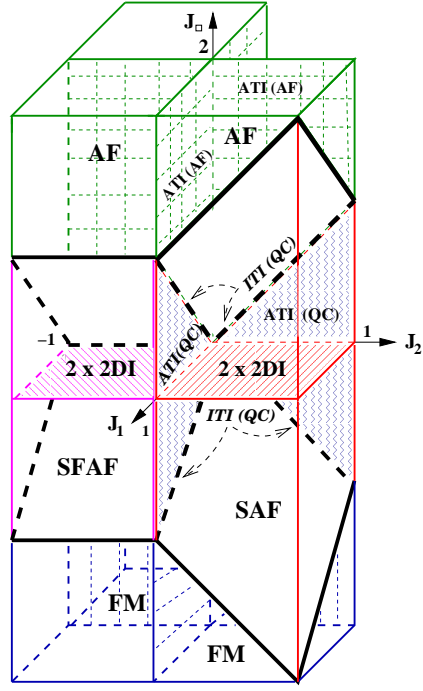}
\caption{Phase diagram of the ground states (GS) of the model
(\ref{NNNIs}). For visualization purposes it is drawn within $(-1,1)
\times (-1,1) \times(-2, 2)$ parallelepiped. The SAF GS (red) lies
between the frustration planes $J_1+J_2=|J_{\text{\tiny
$\square$}}|$ (bold black). The SFAF GS (magenta) lies between the
frustration planes $J_1=|J_{\text{\tiny $\square$}}|$ (bold black).
The AF GS, green (FM GS, blue) lies above (beneath) the frustration
planes and above (beneath) the basal plane in the third quadrant,
respectively. The second quadrant (not shown) is obtained by a
reflection over $J_1=J_2$ plane. Different sectors shown by hatched
and twiggy lines on the exactly-solvable planes $J_{\sharp}=0$ are
explained in the text. } \label{PD3DFul}
\end{figure*}

\begin{figure*}
\includegraphics[width=10.5cm]{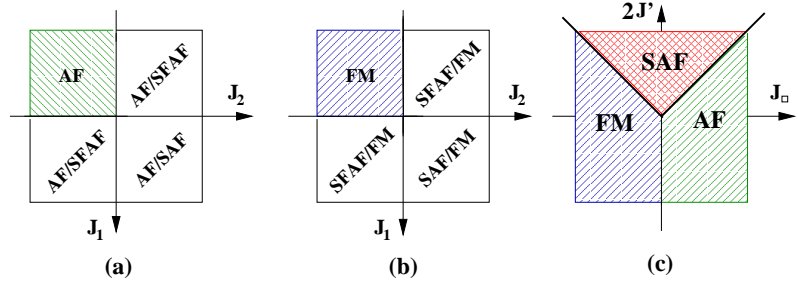}
\caption{Plane projections of the 3D diagram shown in Fig.\
\ref{PD3DFul}. (a): Upper part $J_{\text{\tiny $\square$}}>0$, view
from the top. (b): Lower part $J_{\text{\tiny $\square$}}<0$, view
from the top. (c): Compactification of 3D Fig.\ \ref{PD3DFul} in the
special case $J_1=J_2 \equiv J'$ \cite{FanWu69}. } \label{PDcut}
\end{figure*}

Let us start with the upper part ($J_{\text{\tiny $\square$}}>0$)
of the SAF region
\begin{equation}
\label{SAFup}
J_{\text{\tiny $\square$}}<J_1+J_2
\end{equation}
On the SFAF-SAF boundary $J_2=0$ the model is equivalent to the
anisotropic Ising model on a triangular lattice  (ATI), for which
exact results are available
\cite{Wannier50,Houtappel50,StephensonIII,StephensonIV}. The
antiferromagnetic ($J_{\text{\tiny $\square$}},J_1>0$) ATI model
with one strong bond $J_1>J_{\text{\tiny $\square$}}$ is disordered
at any non-zero temperature \cite{StephensonIV}. It orders only at
$T=0$, i.e., it is Quantum Critical (QC). The highly degenerate
ground state (however with a vanishing zero-temperature entropy per
site) can be viewed as a 2D array of antiferromagnetically ordered
(along the strong bond $J_1$) correlated chains. The oscillating
(with a period of four lattice spacings) power-law decay of the
spin-spin correlation function along $J_{\text{\tiny
$\square$}}$-directions \cite{StephensonIV} indicate on the
preference of the ferromagnetic order along the ``missing" diagonal
$J_2$. This resembles the SFAF state, however any couple of adjacent
$J_1$-chains is uncorrelated. We label this state occurring on two
sectors of the $J_1 (\textrm{or}~J_2)=0$ planes as ATI (QC) on the
phase diagram (Fig.\ \ref{PD3DFul}). Since the critical behavior of
the ATI model with two equal weak ferromagnetic bonds
$|J_{\text{\tiny $\square$}}|<J_1$ is equivalent to the totally
antiferromagnetic ATI model \cite{StephensonIV}, the ATI (QC) state
smoothly continues into the lower ($J_{\text{\tiny $\square$}}<0$)
part of the SFAF-SAF boundary.

The sectors $J_2=0, J_1>0$ (and $1 \leftrightarrow 2$) above the FP
$J_1+J_2=J_{\text{\tiny $\square$}}$ correspond to the
antiferromagnetic ATI model with one weak bond $J_1<J_{\text{\tiny
$\square$}}$. It is known \cite{StephensonIV} to have only two
phases and to order at finite temperature.
$T_c(\textrm{PM}\to\textrm{AF})$ as a function of couplings is also
known exactly. The sectors $J_2=0, J_1<0, J_{\text{\tiny
$\square$}}>0$ (and $1 \leftrightarrow 2$) correspond to the
antiferromagnetic ATI model which is even not  frustrated, and
$T_c(\textrm{PM}\to\textrm{AF})>0$ at any $J_{\text{\tiny
$\square$}}>0$. The PM $\to$ AF transition in the ATI model belongs
to the 2D Ising class. So, except the SFAF-SAF boundary, the ordered
phase on the exactly-solvable ``triangulation" planes $J_1
(\textrm{or}~J_2)=0$ is the same as the AF ground state in the
interior in this region of the phase diagram.

The situation on the ``triangulation" planes in the lower part
($J_{\text{\tiny $\square$}}<0$) of the phase diagram is exactly analogous
to the upper part, with an obvious replacement AF $\leftrightarrow$
FM.

Note that the ground states change on the lines where the
triangulation and frustration planes cross. To put it differently,
these are the lines of quantum phase transitions.
The AF (FM) phase disappears in the limit
$J_1 \to |J_{\text{\tiny $\square$}}|-0~(J_2=0)$ .  Also, the
zero-temperature AF in-chain order [ATI (QC)] described above disappears
in the limit $J_1 \to |J_{\text{\tiny $\square$}}|+0~(J_2=0)$ as well.
The ITI model $J_1 =J_{\text{\tiny $\square$}}$ is disordered at any
non-zero temperature (indicated as ITI (QC) in Fig.\ \ref{PD3DFul}).
Its ground state, albeit having finite entropy per site,
 possesses periodical (with a period of three lattice spacings)
long-range order \cite{StephensonIII}.

The basal plane $J_{\text{\tiny $\square$}}=0$ in Fig.\ \ref{PD3DFul}
corresponds to the case when Hamiltonian (\ref{NNNIs}) represents two
decoupled identical nn Ising models residing on two superimposed
lattices (shown by circles and squares in Fig.\ \ref{IsNNN}). Diagonal
couplings $J_{1,2}$ are the nn couplings of these Ising models.
This is the only exactly-solvable limit (labelled by $2 \times 2$DI
in Fig.\ \ref{PD3DFul}) within the SFAF (or SAF) region of the phase
diagram. In this limit the PM $\to$ SAF (or SFAF) phase
transition enters into the 2D Ising universality class.
%
\subsection{Incommensurate (floating) phase}\label{ICph}
%
So far we have discussed  the ground-state phases of the model and the
critical behavior on the boundaries of these phases with the disordered
phase, as well as on the special planes (lines). However, there is also
a possibility that ordering into the ground-state phases of
Fig.\ \ref{PD3DFul} happens not necessarily from the PM phase, but from
some other one(s) occurring at non-zero temperature. A very simple analysis
indicates that this indeed can take place in our model.
Fourier-transforming the Hamiltonian (\ref{NNNIs}) we obtain (we set
the lattice spacing to unity)
\begin{eqnarray}
\label{HFour}
H &=&  \sum_{\mathbf{q}}
J(\mathbf{q}) T^x(\mathbf{q}) T^x(-\mathbf{q})~, \\ \nonumber
J(\mathbf{q}) &=&
J_{\text{\tiny $\square$}}(\cos q_x + \cos q_y)+
J_1 \cos (q_x -q_y)+J_2 \cos (q_x +q_y)
\end{eqnarray}
where $\mathbf{q}$ runs within the first Brillouin zone $|q_{x,y}|
\leq \pi$. At mean-field level, a minimum of $J(\mathbf{q})$ in
$\mathbf{q}$-space defines the wave-vector $\mathbf{q}_0$ of the
critical freezing mode $T^x(\mathbf{q}_0)$, i.e. the order parameter
$\langle T^x_{\mathbf{m}} \rangle \propto \cos(\mathbf{q}_0
\mathbf{m}+ \varphi)$ below a certain critical temperature $T_c$. In
different regions $(J_1,J_2,J_{\text{\tiny $\square$}})$ of the
ground-state phase diagram (Fig.\ \ref{PD3DFul}) we find minimum at
$\mathbf{q}^{\textrm{\tiny F/A}}=(0,0)/(\pi,\pi)$ giving the FM/AF
order parameter and two minima $\mathbf{q}^{\textrm{\tiny
SAF}}_{1,2}=(\pi,0)/(0,\pi)$ giving two components of the SAF order
parameter. (The latter represent two possible ordering patterns of
the SAF phase and via some transformation can be related to
magnetizations of the superimposed sublattices \cite{KriMu77}). The
important point is that the positions of the \textit{commensurate}
(C) extrema $\mathbf{q}^{\sharp}~(\sharp=$F,A,SAF) of
$J(\mathbf{q})$ do not depend on couplings.

\begin{figure*}
\includegraphics[width=4.5cm]{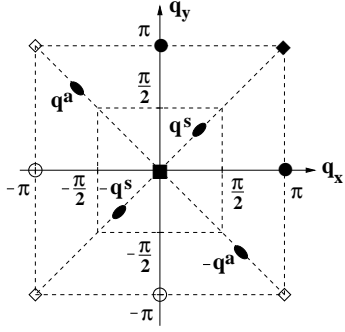}
\caption{Positions of extrema of $J(\mathbf{q})$ (\ref{HFour}) in the
Brillouin zone. Open symbols connected to their bold counterparts by
a reciprocal lattice vector. Incommensurate extrema
$\pm \mathbf{q}^{a,s}$ (\ref{ICexts},\ref{ICexta}) (shown for the case
$J_1>0,J_2<0$) exist if conditions (\ref{ICcond}) are satisfied.
}
\label{Bril}
\end{figure*}

There are however two other pairs of extrema $\pm \mathbf{q}^{s,a}$
which exist if
\begin{equation}
\label{ICcond}
|J_{\text{\tiny $\square$}}| <2|J_1| \quad \textrm{and/or} \quad 2|J_2|
\end{equation}
These extrema lie on the diagonal of the Brillouin zone.
$\mathbf{q}^{s,a}$ are generically \textit{incommensurate} (IC)
and depend on couplings as
\begin{eqnarray}
\label{ICexts}
q_{x}^s &=&\phantom{-} q_{y}^s =  \arccos
\big ( -\frac{J_{\text{\tiny $\square$}} }{2J_2} \big ) \\
\label{ICexta}
q_{x}^a &=&-q_{y}^a =  \arccos
\big ( -\frac{J_{\text{\tiny $\square$}} }{2J_1} \big )
\end{eqnarray}
We show the positions of all extrema of $J(\mathbf{q})$ within the
Brillouin zone in Fig.\ \ref{Bril}. In the SFAF ground state region
(e.g., in the  fourth quadrant $J_1>0,J_2<0$ shown in Fig.\
\ref{PD3DFul}) the pair of  extrema $\pm \mathbf{q}^a$
(\ref{ICexta}) gives global minima of $J(\mathbf{q})$ (the other
pair of solutions (\ref{ICexta}) $\pm \mathbf{q}^s$ when exists,
corresponds to its maxima), and $|q_{x}^a| =|q_{y}^a| >\pi/2$. As we
see, two IC modes $\pm \mathbf{q}^a$ could give components of the
SFAF ground state order parameter's wave vectors $\pm
\mathbf{q}^{\textrm{\tiny SFAF}}= \pm (\pi/2, -\pi/2)$ only if
$J_{\text{\tiny $\square$}}=0$. (In the second quadrant of the SFAF
region when $J_1<0,J_2>0$, the vectors $\mathbf{q}^s$ and
$\mathbf{q}^a$ exchange their roles. Because of the $J_1
\leftrightarrow J_2$ symmetry, in the following we will always
discuss the fourth quadrant for concreteness.)

\begin{figure*}
\includegraphics[width=4.5cm]{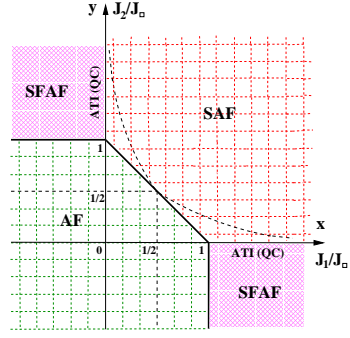}
\caption{Plane phase diagram for the ratios of the couplings,
corresponding to the 3D Fig.\ \ref{PD3DFul} at
$J_{\text{\tiny $\square$}}>0$. The phase boundaries (thick solid lines)
corresspond to the frustrations planes FP and FP'. The thick dashed lines
(black, online) indicate the boundaries of the incommensurate global
minima locus ($y>1/2,~x>1/2, ~y<1/4x$) discussed in the text. The case
$J_{\text{\tiny $\square$}}<0$ can be  obtained
by the substitutions $J_{1,2}/J_{\text{\tiny $\square$}} \mapsto
J_{1,2}/|J_{\text{\tiny $\square$}}|$, AF $\mapsto$ FM in this figure.
}
\label{PDpl}
\end{figure*}

The locus of the IC global minima  does not coincide with the SFAF
ground state region, but overlaps with the neighboring FM, AF, and
SAF phases. The  minimum $J(\pm \mathbf{q}^a)$ is located between
two planes $|J_{\text{\tiny $\square$}}|= 2J_1$ in the fourth
quadrant, and in two regions of the half of the first quadrant
($J_2<J_1$): (i) between $J_{\text{\tiny $\square$}} = +2J_1$ and
$J_{\text{\tiny $\square$}} = +2 \sqrt{J_1 J_2}$; (ii) $+ \mapsto
-$. The regions of the IC minima in the other half ($J_2>J_1$) of
the first quadrant (as in the second quadrant) are obtained from the
described above by $J_1 \leftrightarrow J_2,~\mathbf{q}^a \mapsto
\mathbf{q}^s$. On the plane phase diagram shown in Fig.\ \ref{PDpl}
this locus is restricted by the lines $y=1/2,~x=1/2,~y=1/4x$, shown
by the black dashed lines.

So we can conclude that at finite temperature the model possesses an
IC phase and there is an IC-C phase transition where the IC wave
vector $\mathbf{q}^a$ locks into one of the (commensurate)
ground-state phase vectors. As in 2D the IC phase has only an
\textit{algebraic} long-range order, it is called floating
\cite{Bak82}. The origin of the IC floating phase in our model is
frustration (competing interactions). Such phase is well known from
another example of frustrated Ising model, i.e., the ANNNI model
which was intensively studied in the past
\cite{Liebmann86,Bak82,Selke88}. In that model the floating phase
locks into the antiphase which has the wave vector
$\mathbf{q}=(0,\pi/2)$. (The antiphase is analogue of our SFAF
phase.) The ANNNI model also provides an example showing that the
mean-field (minimization) analysis does not work well in defining
boundaries between the floating and commensurate phases in 2D, and
the extend of the IC phase is \textit{less} than the mean field
suggests \footnote{ \label{ANNNIcom} The well-studied ANNNI phase
diagram with the FM and antiphase ground states
\cite{Liebmann86,Bak82,Selke88} is an analogue of the lower part of
the fourth quadrant (FM-SFAF) of our diagram in Fig.\
\ref{PD3DFul}}.

We will not attempt to locate exactly the phase boundaries at finite
temperature in this study. Following Domany \textit{et al}
\cite{Dom78} in classification of the ordered phases by
commensurability $p$, i.e., the ratio of superstructure's period and
lattice spacing along a given direction, we can label the phases as
follows: F $\mapsto~1 \times 1$; AF $\mapsto~2 \times 2$; SAF
$\mapsto~1 \times 2$ (or $2 \times 1$); SFAF $\mapsto~4 \times
4$\footnote{ \label{noteAnti} According to this notation, the
antiphase of the 2D ANNNI model is $(1 \times 4)$. For the same
reasons we give for our case, the floating phase of that model
exists only within the antiphase ground state region.}. From mapping
of the 2D IC-C phase transition to the Kosterlitz-Thouless problem,
it is established that there is no such (continuous) transition for
commensurate phases with small $p^2 <8$ \cite{Bak82,Copper82/1}.
From this result with the proviso of continuity of phase
transition(s) in the model, we conclude that the IC floating phase
cannot ``spill" beyond the SFAF ($p=4$) ground-state region of the
phase diagram, even if a naive (mean-field) analysis suggests that
within the F, AF, and SAF regions there are some parts where it
could be possible (see Fig.\ \ref{PDpl}). The only high-temperature
phase the latter three regions have a common border, is the
disordered PM.

Combining this with the known exact critical properties of the model
on the special planes discussed above, we end up with the
qualitative finite-temperature phase diagram shown in Fig.\
\ref{Float}. Since on the plane $J_{\text{\tiny $\square$}}=0$ the
model (\ref{NNNIs}) is just two decoupled Ising lattices, the
floating phase must be absent. Mean-field arguments suggest that the
floating phase disappears exactly at $J_{\text{\tiny $\square$}}=0$
giving rise to a Lifshitz point (L). Note that the floating phase
does not appear if the diagonal couplings are equal even at the
mean-field level (see Fig.\ \ref{PDpl}), which agrees with known
more sophisticated analyses of this case \cite{Liebmann86}.
%
\subsection{The model in transverse field}\label{TrF}
\begin{figure*}
\includegraphics[width=9.5cm]{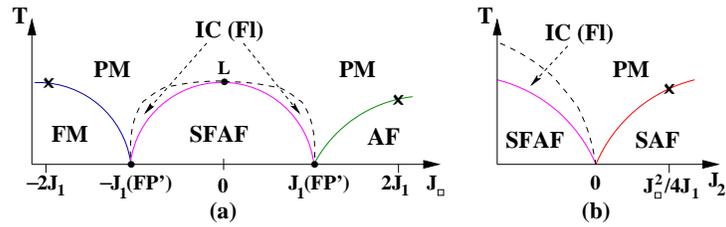}
\caption{Qualitative temperature phase diagram. The IC floating (Fl)
phase lies beneath the dashed lines. Crosses indicate the borders of
the IC minima locus. (a): For fixed $J_1>0,~J_2<0$. On the plane
$J_{\text{\tiny $\square$}}=0$ the floating phase must be absent. We
assume that it smoothly disappears at $J_{\text{\tiny $\square$}}=0$
resulting in a Lifshitz point (L). (b): The same for fixed
$J_1>0,~|J_{\text{\tiny $\square$}}| <J_1$. } \label{Float}
\end{figure*}

%
Now we turn to the analysis of the nn and nnn Ising Hamiltonian $H$
(\ref{NNNIs}) in the presence of a transverse field. The total Hamiltonian
of the Ising Model in Transverse Field (IMTF) reads
\begin{equation}
\label{IMTF1}
H_{\textrm{IMTF}}= H - \Omega \sum_{\mathbf{i}} \mathcal T^z_{\mathbf{i}}
\end{equation}
The Ising operators are normalized to satisfy the spin algebra
\begin{equation}
\label{SpinAl}
 \big[ \mathcal T^{\alpha}_{\mathbf{i}},\mathcal T^{\beta}_{\mathbf{j}}
 \big]= i \delta_{\mathbf{ij}} \epsilon_{\alpha \beta \gamma}
\mathcal T^{\gamma}_{\mathbf{i}}
\end{equation}
There is no exact solution of the transverse 2D Ising model even for
the case of nn couplings only ($J_1=J_2=0$). The ground state phase
diagram of the Hamiltonian (\ref{IMTF1}) can be analyzed from the
known mapping of the $d$-dimensional IMTF at zero temperature onto
the $(d+1)$-dimensional Ising model at a given (non-zero)
``temperature" \cite{Chak96}. In our case the 2D nn and nnn IMTF
maps onto the 3D Ising model comprised of the 2D layers
(\ref{NNNIs}) coupled \textit{ferromagnetically} in the third
(Trotter) direction with the coupling $J_{T} \propto -\ln \coth
\Omega<0$. For such $(2+1)$-dimensional model a mean-field analysis
gives a qualitatively correct diagram of the \textit{ground state}
phases of the 2D IMTF \cite{Chak96}. The new coupling $J_{T}$ does
not bring any additional frustration to the 2D nn and nnn model.
Analysis of $J_3(q_x,q_y,q_T)=J_T \cos q_T +J(q_x,q_y)$ where
$J(q_x,q_y)$ is given by (\ref{HFour}), shows that $J_T$ does not
modify the domains of the global minima in the
$(J_1,J_2,J_{\text{\tiny $\square$}})$-space, adding only a trivial
$q_T=0$ third component to the two-dimensional vectors
$\mathbf{q}^{\sharp}$ discussed above (cf. Fig.\ \ref{Bril}.) The
temperature phase diagram of the 3D Ising model with the spectrum
$J_3(q_x,q_y,q_T)$ (if we label the phases according to the in-plane
ordering pattern defined by the 2D vectors $\mathbf{q}^{\sharp}$)
looks similar to the one shown in Fig.\ \ref{Float}, with one very
important distinction: the abovementioned argument related to
phase's commensurability $p$ does not apply in 3D, so the IC region
is not restricted to lie above the SFAF phase, but can spill into
the neighboring regions of the $(J_1,J_2,J_{\text{\tiny
$\square$}})$-space. From the mean-field arguments, the IC region is
given by the locus of the IC minima $J(\pm \mathbf{q}^{a/s})$
defined in the previous section. So, the IC phase instead of being
locked between the special planes FP' and $J_2=0$ as shown in Fig.\
\ref{Float} (a) and (b), respectively, can spread up to the locus
boundaries shown by the crosses. From the equivalence between the
zero-temperature $d$-dimensional IMTF (quantum Ising) and the
$(d+1)$-dimensional classical Ising model, we infer that the
ground-state phase diagram of the former on the plane ($\Omega/J,
J_{\sharp}$) should have the same structure as the described above
($T, J_{\sharp}$) diagram of the latter. So we expect the transverse
field to generate the IC ground-state phase not only in the SFAF
region of the the Ising coupling space, but also in the neighboring
parts of the F, AF, and SAF regions. From minimization arguments the
latter are restricted by the dashed lines on the plane diagram in
Fig.\ \ref{PDpl} and by the crosses in Fig.\ \ref{Float}.

From analogies with the ANNNI model, we rather expect this ``IC
region" to be filled with infinitely many commensurate phases with
different $p$ \cite{Bak82,Selke88}, but detailed analysis of this
question, as well as the full \textit{finite-temperature} phase
diagram of the transverse model (\ref{IMTF1}) need a separate study.

In the rest of the paper we will be  particularly interested in the SAF
region of the coupling space, and restrict ourselves to the couplings
\begin{equation}
\label{SAFsafe}
J_{\text{\tiny $\square$}}^2 <4 J_1 J_2, \quad
(J_1,J_2)>0
\end{equation}
According to Fig.\ \ref{PDpl}, it means that we choose the couplings
to lie above the hyperbole $y=1/4x$. The first condition in
(\ref{SAFsafe}) ensures that the couplings lie in the region where
$\mathbf{q}^{\textrm{\tiny SAF}}_{1,2}=(0,\pi)/(\pi,0)$ provide a
global minimum of model's spectrum, so the phase with the IC
solutions $\mathbf{q}^{a/s}$ does not intervene. The second in the
above conditions stipulates that even if $J_{\text{\tiny
$\square$}}=0$ we stay away from the planes where our model becomes
the triangular Ising. A transverse field can can generate exotic
temperature phases in that model. Such phases were found
\cite{Moessner01,Most03} in the particular case of the isotropic
(antiferromagnetic) transverse triangular Ising model\footnote{
\label{noteDeg} By analogy with the triangular Ising case of
Refs.[\onlinecite{Moessner01,Most03}], we expect the transverse
field to bring about new exotic phases near the frustration planes
FP, FP', where the ground state of our model is also infinitely
degenerate.}.

From mapping between the quantum and classical Ising models we
conclude that at zero temperature our IMTF with the couplings
satisfying (\ref{SAFsafe}) possesses a single Quantum Critical Point
(QCP) which separates the SAF and PM ground state phases. The mean
field predicts a two-phase PM/SAF diagram. The critical temperature
$T_c$ of the second-order PM-SAF phase transition evolves smoothly
from the QCP $T_c(J/ \Omega_{\textrm{cr}})=0$ to the asymptotic
limit $T_c(\infty)$ of the classical model (see dashed curve in
Fig.\ \ref{TcFig}, where $g \equiv (J_1+J_2)/ \Omega$). It is also
known that the mean field gives a qualitatively correct phase
diagram for the IMTF when $d \geq 2$ \cite{Chak96}. Thus we argue
that the mean-field result shown by the dashed curve in Fig.\
\ref{TcFig} \textit{does represent} the phase diagram of the IMTF
(\ref{NNNIs},\ref{IMTF1},\ref{SAFsafe}), while its quantitative
aspects, e.g., the exact value of the QCP, should be corrected via
more accurate treatments.
%
\section{Coupled Spin-Pseudospin Model}\label{SPM}
%
\subsection{Hamiltonian}\label{STHam}
%
Now we turn to the analysis of the IMTF (\ref{NNNIs},\ref{IMTF1})
coupled to the quantum spins ($\mathbf{S}$) residing on the same
sites of the lattice as the Ising spins do. The latter we will call
\textit{pseudospins} from now on. Such coupled spin-pseudospin (or
spin-orbital) models emerge in various contexts, most notably the
Jahn-Teller transition-metal compounds \cite{Kugel82} or many kinds
of low-dimensional quantum magnets. For a recent short overview and
more references, see \cite{CGbig03}.

Models of the type we study in the present work appear in analyses
of the quarter-filled ladder compound $\rm NaV_2O_5$. The
Hamiltonian of this material can be mapped onto a spin-pseudospin
model with spins and pseudospins residing on the same rung of a
ladder \cite{Most98,Sa00,Most02}. The ladders form a 2D lattice. The
long-range pseudospin order $\langle T^x_{\mathbf{i}} \rangle \neq
0$ represents physically the charge disproportionation between
left/right sites on a rung below $T_c$.

This system was analyzed on the effective triangular lattice shown
in Fig.\ \ref{LatMapFig}a by solid lines \cite{Most98,Most02}.
However, in the case of $\rm NaV_2O_5$ the Ising couplings generated
by Coulomb repulsion are antiferromagnetic and $J_1>J_{\text{\tiny
$\square$}}$. Since the triangular Ising model with one strong side
is disordered \cite{StephensonIV}, one needs an extra coupling
($J_2$-diagonal) to stabilize the observed SAF long-range order. In
our earlier study \cite{CGbig03} we explicitly took into account
$J_1$, while the other diagonal $J_2$ was effectively generated via
the spin-pseudospin coupling\footnote{ \label{effCoup} A linear
coupling $\propto \varphi (\mathcal T_1 \mp \mathcal T_2)$ with some
Gaussian mode $\varphi$ results in an effective
anti-/ferrro-magnetic interaction between $\mathcal T_1$ and
$\mathcal T_1$. Such terms can create or renormalize the couplings
of the Ising effective Hamiltonian. In his context the Ising model
with two nnn couplings of different signs is less academic than it
might appear.}.

In this work we take into account the Ising couplings
$J_{\text{\tiny $\square$}},J_1,J_2$ between neighboring sites of
the effective lattice, as shown in Fig.\ \ref{LatMapFig}a. Then such
effective lattice can be mapped onto the square lattice shown in
Fig.\ \ref{LatMapFig}b with the nn and nnn Ising couplings. For $\rm
NaV_2O_5$ all $J_{\sharp} >0$, so the model is frustrated. As
follows from geometry of the original $\rm NaV_2O_5$ lattice, $J_2$
is the weak diagonal and $J_1$ is the largest coupling:
\begin{equation}
\label{CoupCond}
J_2<J_{\text{\tiny $\square$}} <J_1
\end{equation}
We assume $J_{\text{\tiny $\square$}}$ to be small enough not only
to lie beneath the frustration plane (\ref{FP}), but to satisfy a
more stringent condition (\ref{SAFsafe}). Then according to the
above analysis, the IMTF with these couplings has a two-phase
(PE-SAF) diagram with a QCP\footnote{ \label{noteEl} Since in
applications to $\rm NaV_2O_5$ the Ising pseudospins $\mathcal T$
represent charge displacements, the appropriate names for the phases
are ``paraelectric" (PE) and ``super-antiferrolelectric". We keep
the same abbreviation SAF for the latter.}. In the literature on
$\rm NaV_2O_5$ its charge order is called the ``zig-zag phase", what
characterizes the antiferroelectric order in a \textit{single
ladder} only. In fact, the \textit{two-dimensional long-range charge
order} in $\rm NaV_2O_5$ is SAF. For a detailed explanation of this
point, including interpretation of the experimental crystallographic
data on the charge order \cite{Grenier02} in terms of the Ising
pseudospins, see \cite{Stack04}.

In the following we will work with dimensionless quantities:
Hamiltonians ${\mathcal H}=H/ \Omega$, temperature $T \rightarrow
T/\Omega$ and Ising couplings $g_{\sharp} \equiv J_{\sharp}/
\Omega$. With site labelling shown in Fig.\ \ref{LatMapFig}, the
IMTF Hamiltonian is:
\begin{equation}
\label{HIMTFdls} \mathcal{H}_{\textrm{IMTF}}=\sum_{m,n} \Big\{ -
\mathcal T^z_{mn}+ \frac12  \Big[ g_{\text{\tiny $\square$}}
\Big(\mathcal T^x_{mn} \mathcal T^x_{m+1,n+1}+ \mathcal T^x_{mn}
\mathcal T^x_{m+1,n} \Big) + g_1 \mathcal T^x_{mn} \mathcal
T^x_{m,n+1} + g_2 \mathcal T^x_{mn} \mathcal T^x_{m+2,n+1} \Big]
\Big\}
\end{equation}

For the decoupled spin sector of the total Hamiltonian we take into
account only the strongest coupling between spins on the nn rungs of
a ladder. In terms of the effective lattice (cf. Fig.\
\ref{LatMapFig}) this translates into a set of decoupled Heisenberg
chains with the usual antiferromagnetic spin exchange $J$. These
parallel chains are oriented along the $J_1$-diagonals.

As we infer from our previous work on a simpler version of the IMTF
Hamiltonian \cite{CGbig03}, there are two spin-pseudospin
interaction terms resulting in two qualitatively distinct aspects of
model's criticality: the inter-ladder spin-pseudospin interaction
$\propto \varepsilon$, and the in-ladder spin-pseudospin interaction
$\propto \lambda$. The former, in terms of the equivalent square
lattice, linear over difference of the charge displacement operators
$\mathcal T^x$ on the nnn sites along the weak $J_2$-diagonal, is
responsible for the simultaneous appearance of the SAF order and the
spin gap, as well as for the destruction of the IMTF QCP\footnote{
\label{noteSym} This type of coupling is allowed by the symmetry of
the original $\rm NaV_2O_5$ lattice \cite{CGbig03}. Numerical
estimates of $\varepsilon$ from a microscopic Hamiltonian are given
in \cite{Diag05}. Note also that $\varepsilon$ effectively couples
the spin chains.}. The latter, quadratic over the nnn charge
operators along the $J_1$-diagonal, is responsible for the
re-entrance. With all these terms the total effective Hamiltonian
reads:
\begin{eqnarray}
\label{Hamdls} \mathcal H= \mathcal H_{\textrm{IMTF}} +  \sum_{m,n}
{\textbf S}_{mn}{\textbf S}_{m,n+1} \big[ J +\lambda \mathcal
T^z_{mn} \mathcal T^z_{m,n+1}+ \varepsilon \big( \mathcal
T^x_{m+1,n+1}- \mathcal T^x_{m-1,n}\big) \big]
\end{eqnarray}
The dimensionless couplings $J,\lambda,\varepsilon$ in the
Hamiltonian (\ref{Hamdls}) are positive, and the spin operators
satisfy the same algebra (\ref{SpinAl}) as the pseudospins (while
$S$ and $\mathcal T$ commute). The sums above run through $1 \leq m
\leq \mathcal M$ and $1 \leq n \leq \mathcal N$. For brevity we will
use the notation $D_{mn} \equiv {\textbf S}_{mn} {\textbf
S}_{m,n+1}$. In this  study we consider the model with the $XY$ spin
sector:
\begin{equation}
\label{Sint}
D_{mn}=
S_{mn}^x S_{m,n+1}^x+S_{mn}^y S_{m,n+1}^y
\end{equation}
The range of couplings under consideration will be restricted to
\begin{equation}
\label{parange}
(J, \lambda) \alt  g_{1}, \quad
\varepsilon \alt \textrm{max} \{J, \lambda \}
\end{equation}
%
\subsection{Spin-SAF phase transition}\label{PhTr}
%
%
We treat the Hamiltonian (\ref{Hamdls}) following conventional
wisdom of molecular-field approximations (MFA) \cite{Blinc74}. In
the present version of MFA the pseudospins are decoupled and
averaged with the density matrix $\rho^{\mathcal T} \propto
\textrm{exp}(-\beta \textbf h_{mn} \bm{\mathcal T}_{mn})$, where
$\textbf h_{mn}$ is the Weiss (molecular) field, while the spin
sector is treated exactly via a Jordan-Wigner transformation. The
details are presented in \cite{CGbig03}. Similar to the pure IMTF
with couplings (\ref{SAFsafe}) we assume the possibility of the SAF
order in the coupled model (\ref{Hamdls}). So we take the following
Ans\"atze for the Ising pseudospin averages (i.e., the charge
ordering parameters in terms of the real physical quantities)
\begin{eqnarray}
\label{mz}
\langle \mathcal T_{mn}^z \rangle &=& m_z \\
\label{mx}
\langle \mathcal T_{mn}^x \rangle &=&(-1)^{m+n}m_x
\end{eqnarray}
It is easy to see from the Hamiltonian (\ref{Hamdls}) that ansatz
(\ref{mx}) creates a dimerization in the spin sector, therefore a natural
assumption for the dimerization operator average is
\begin{equation}
\label{Dmn}
\langle D_{mn} \rangle =-[t+(-1)^{m+n} \delta]
\end{equation}
With the new coupling
\begin{equation}
\label{g}
g \equiv g_1+g_2
\end{equation}
the molecular-field equations and results derived in \cite{CGbig03}
for the case $g_2=g_{\text{\tiny $\square$}} =0$ (i.e., $g=g_1$) can
be applied here. Some of them we reproduce in this paper in order to
make it more self-contained, and for the use in what follows as
well.

The average quantities are determined by the system of four
coupled equations
\begin{subequations}
\label{EqMF}
\begin{eqnarray}
m_z
&=&\frac12
\frac{1+2\lambda t m_z}{\mathfrak h} \tanh\frac{\beta \mathfrak h}{2} \\
m_x
&=&\frac{m_x}{2}
\frac{g+2\varepsilon \eta}{\mathfrak h} \tanh\frac{\beta \mathfrak h}{2} \\
t &=& \frac{1}{\pi} \int_0^{\frac{\pi}{2}}
d \varphi \frac{\cos^2 \varphi}{\xi(\varphi)}
\tanh  \tilde \beta \xi(\varphi)
\equiv \frac{1}{\pi}t_n(\Delta, \tilde \beta) \\
\eta &=& \frac{\Delta}{\pi m_x} \int_0^{\frac{\pi}{2}}
d \varphi \frac{\sin^2 \varphi}{\xi(\varphi)}
\tanh \tilde \beta \xi(\varphi)
\equiv \frac{\Delta}{\pi m_x} \eta_n(\Delta, \tilde \beta)
\end{eqnarray}
\end{subequations}
where $\mathfrak h =\sqrt{h_x^2+h_z^2}$ is the absolute value of the Ising
molecular field
\begin{subequations}
\label{MolF}
\begin{eqnarray}
h_z &=& 1+ 2\lambda t m_z \\
h_x &=& g m_x+ 2\varepsilon \delta
\end{eqnarray}
\end{subequations}
The other auxiliary parameters are defined as follows:
\begin{subequations}
\label{Param}
\begin{eqnarray}
\delta &\equiv&  m_x \eta \\
\xi(\varphi) &\equiv& \sqrt{ \cos^2 \varphi+\Delta^2 \sin^2 \varphi  } \\
\Delta       &\equiv& \frac{2 \varepsilon m_x}{J+\lambda m_z^2} \\
\tilde \beta   &\equiv& \frac{\beta}{2}(J+\lambda m_z^2)
\end{eqnarray}
\end{subequations}
At some critical temperature $T_c$ the coupled model undergoes the
phase transition. It is of the second kind, with the thermodynamic
behavior of the physical quantities as the Landau theory of phase
transitions predicts \cite{CGbig03}. With the spin-pseudospin
(-charge) coupling $\varepsilon$ present, the SAF charge order $m_x
\neq 0$ and the spin gap $\Delta_{\textrm{SG}} = 2 \varepsilon m_x$
appear simultaneously below $T_c$. By analogy with the spin-Peierls
transition, when the Peierls phonon instability (freezing) creates
the spin gap, it is natural to call this type of transition the
\textit{spin-super-antiferroelectric (spin-SAF)} transition.

It is worth to point out an important property of the Hamiltonian
(\ref{Hamdls}): in the other domains of Ising couplings (not considered in
our analysis of the coupled model) where the Ising sector of (\ref{Hamdls})
can order into, e.g., FM, AF, or SFAF phase, the dimerization (gap) in the
spin sector does not occur.

The behavior of $T_c(g)$ in the coupled model (\ref{Hamdls}) shows
two new striking features comparatively to the pure IMTF:
re-entrance and destruction of the QCP \cite{CGbig03}. In the
absence of the spin-charge coupling $\varepsilon$, the model
(\ref{Hamdls}) has a QCP at
\begin{equation}
\label{glam}
g_{\lambda} \equiv 2 \Big(1+ \frac{\lambda}{\pi} \Big),
\end{equation}
where $T_c$ vanishes (see Fig.\ \ref{TcFig}). $\lambda$ renormalizes
the QCP comparatively to the pure IMTF value $g=2$. The coupling
$\varepsilon$, responsible for the spin gap generation also destroys
the QCP, resulting in the exponential behavior of $T_c$ in the
region $g \lesssim g_{\lambda}$, where the model would have been
disordered at any temperature if $\varepsilon=0$. This constitutes
an important feedback from the spins on the charge degrees of
freedom, allowing the very possibility of the model \textit{to order
at all.} Approximate analytical solutions for $T_c(g)$ in the
regimes of strong Ising couplings and the BCS- are: \cite{CGbig03}
\begin{equation}
\label{Tcas}
 T_c \approx  \left\{
                \begin{array}{ll}
  \frac{g}{4}, & g \gg g_{\lambda} \\[0.2cm]
 \frac{\mathbb A \tilde J }{2} \textrm{exp}
 \big[- \frac{\pi \tilde J}{4 \varepsilon^2}(g_{\lambda}-g) \big]
  , &\textrm{BCS regime}
                \end{array}
       \right.
\end{equation}
where $\mathbb A \equiv\frac{8}{\pi \textrm{e}^{1- \gamma}} \approx 1.6685$,
$\gamma \approx 0.5772$ is Euler's constant, and
\begin{equation}
\label{Jbar}
\tilde J \equiv J +\frac{\lambda}{4}
\end{equation}
The boundary where the low-temperature BCS regime sets in and the related
formulas are applicable, is given approximately by the condition
\begin{equation}
\label{BCScond}
\textrm{BCS regime}:~
g < g_{\lambda} +\frac{4 \varepsilon^2}{\pi \tilde J}
\end{equation}
The BCS regime has many analogies with the standard theory of
superconductivity, apart from the exponential dependence of $T_c$ on
couplings. In particular, several physical quantities (order
parameter, BCS ratio, specific heat jump) manifest certain
``universal" behavior near $T_c$, similar to that known from the BCS
theory \cite{CGbig03}.

Another particularity of $T_c(g)$ found earlier \cite{CGbig03} from
the numerical solution of Eqs.(\ref{EqMF}), is re-entrance in the
intermediate regime $g \sim g_{\lambda}$. The re-entrance occurs in
the coupled model with the QCP ($\varepsilon=0$), while when
$\varepsilon \neq 0$ the critical temperature can even manifest a
double re-entrant behavior before it reaches the BCS regime (see
Fig.\ \ref{TcFig}). A detailed analysis of the coupled model in the
regime of re-entrance was not done previously. We will address this
in the next subsection, mainly analytically, in order to get more
insight on the underlying physics and, in particular, to establish
conditions when the re-entrance can occur.
%
\subsection{Re-entrance}\label{Reen}
%
%
%
Let us first reproduce some earlier formulas \cite{CGbig03} for
reader's convenience. At $T \leq T_c$ one equation from the pair
(\ref{EqMF}a,b) can be written in a form
\begin{equation}
\label{mzbe}
m_z^{-1} = g+ \frac{4 \varepsilon^2}{\pi(J+\lambda m_z^2)} \eta_n
 -\frac{2\lambda }{\pi} t_n , ~~T \leq T_c
\end{equation}
At $T=T_c$ we have Eqs.(\ref{EqMF}a) as
\begin{equation}
\label{mzTc}
m_z = \frac12 \tanh\frac{\beta_c }{2} (1+\frac{2\lambda m_z}{\pi}t_n)~,
\end{equation}
and parameters $t_n,\eta_n$ are given by Eqs.(\ref{EqMF}c,d) with
$\Delta=0$. The latter two functions have the following expansions:
\cite{CGbig03}
\begin{equation}
\label{tnas}
t_n(0,x) \approx  \left\{
                \begin{array}{ll}
 \frac{\pi}{4} x(1-\frac14 x^2) + \mathcal O (x^5 ) , & x <1\\[0.2cm]
  1-\frac{\pi^2}{24} \frac{1}{x^2}+
  \mathcal O(\frac{1}{x^4}), & x >1
                \end{array}
       \right.
\end{equation}
and
\begin{equation}
\label{etanas}
\eta_n(0,x) \approx \left\{
                \begin{array}{ll}
  \frac{\pi}{4} x(1-\frac{1}{12}x^2) + \mathcal O(x^5 ) , & x <1 \\[0.2cm]
 \ln \mathbb A x+\frac{\pi^2}{48} \frac{1}{x^2}+
 \mathcal O(\frac{1}{x^4}), & x >1
                \end{array}
       \right.
\end{equation}
%
\subsubsection{Case $\varepsilon=0$; re-entrance with QCP}\label{eps0}
%
As one can easily see from Eqs.(\ref{mzbe},\ref{mzTc}) there is no
re-entrance when $\lambda=0$. This is a well-known fact for the pure
IMTF, as on the mean-field level, as well as beyond MFA
\cite{Blinc74,Chak96}. To study the re-entrance analytically and in
particular, to establish whether there is some minimal value of
$\lambda$ when it appears, we should distinguish between two
asymptotic regimes of the mean-field equations. Let us first
consider the regime (it can occur only if $\tilde J <1$) when ($T_c
\equiv 1/ \beta_c$)
\begin{equation}
\label{tilbetsmall}
\frac12 \tilde J <T_c< \frac12
\end{equation}
(In all regimes of couplings the re-entrance occurs at
$T_c<\frac12$). By carrying out the leading-term expansions of the
functions in Eqs.(\ref{mzbe},\ref{mzTc}) we obtain
\begin{equation}
\label{gTcsmall}
g = 2+4 e^{-1/T_c} +\frac{\lambda \tilde J}{4 T_c}
\end{equation}
for a single-valued function $g(T_c)$. The non-monotonic (i.e.,
re-entrant) behavior of $T_c(g)$ is related to the existence of an
extremum of $g(T_c)$. The coupling $g_{\textrm{min}}$ which defines the
minimal value of $g$ for the order in $m_x$ being possible (in the pure
IMTF with $\lambda=0$ this was the QCP), and in the
same time the left border of the re-entrant region
\begin{equation}
\label{reentrI}
g_{\textrm{min}} <g<g_{\lambda},
\end{equation}
is defined from the  minimum of the function $g(T_c)$
(\ref{gTcsmall}). This point corresponds to the critical temperature
\begin{equation}
\label{Tstar1}
T_{*}=\ln^{-1} \kappa_{\circ}, ~
\kappa_{\circ} \equiv \frac{16}{\lambda \tilde J}
\end{equation}
for which
\begin{equation}
\label{gmin1}
g_{\textrm{min}} \approx 2+ \frac{4}{\kappa_{\circ} }
\big(1+\ln\kappa_{\circ} \big)
\end{equation}
The consistency of the solution (\ref{Tstar1}) with
(\ref{tilbetsmall}) implies the condition
\begin{equation}
\label{Condri}
2 < \ln\kappa_{\circ}
<\frac{2}{ \tilde J}~,
\end{equation}
The other regime corresponds to the case when
\begin{equation}
\label{tilbetlarge}
T_c< \textrm{min} \{1/2, \tilde J/2 \}
\end{equation}
Proceeding in the same way as above, we obtain for $g(T_c)$ in this
case:
\begin{equation}
\label{gTclarge}
g =g_{\lambda}+4 e^{-g_{\lambda}/2T_c} -
\frac{\lambda \pi}{3 \tilde J^2} T_c^2
\end{equation}
Let us point out that the conditions
(\ref{tilbetsmall},\ref{tilbetlarge}) determine two different
regimes ($x<1$ or $x>1$) of the asymptotics
(\ref{tnas},\ref{etanas}) we apply in order to obtain $g(T_c)$ as
(\ref{gTcsmall}) or (\ref{gTclarge}). So if $\tilde J<1$ there are
regions of $T_c$ where condition (\ref{tilbetsmall}) is satisfied,
then the approximation (\ref{gTcsmall}) applies. However at
sufficiently low temperatures ($T_c<\tilde J/2$) we inevitably enter
the other regime (\ref{tilbetlarge}) where the asymptotics
(\ref{tnas},\ref{etanas}) change ($x<1~\mapsto~x>1$), and the
function $g(T_c)$ crosses over from (\ref{gTcsmall}) to
(\ref{gTclarge}). If, on the contrary, $\tilde J$ is large, then
condition (\ref{tilbetsmall}) never applies, and the approximation
(\ref{gTclarge}) describes the whole region $T_c<1/2$.

\begin{figure*}
\includegraphics[width=10.0cm]{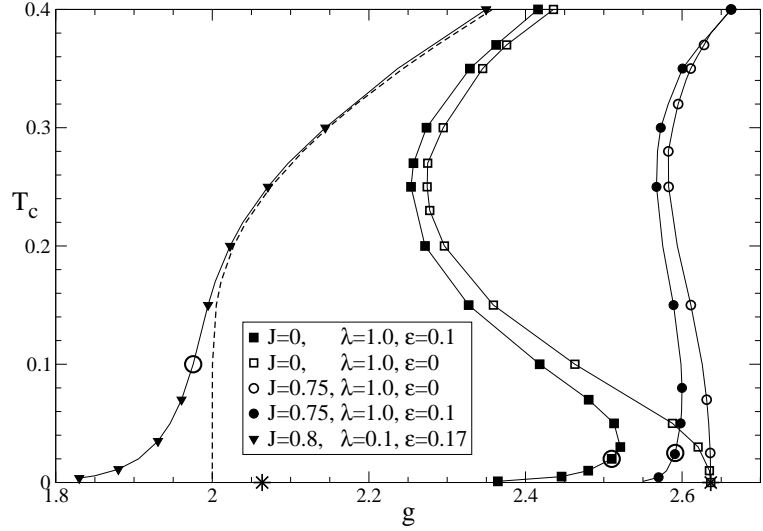}
\caption{
Critical temperature of the PE-SAF phase transition as a function of the
Ising coupling $g$ at different values of $J,\lambda, \varepsilon$
from the numerical solution of Eqs.(\ref{EqMF}). The
dashed line corresponds to the pure IMTF ($J=\lambda=\varepsilon=0$).
Two stars on the abscissa show the positions of critical couplings
$g_{\lambda}=2.0627~(2.6366)$ for $\lambda=0.1~(1.0)$, resp. Large empty
circles on the curves with $\varepsilon \neq 0$ indicate the right
boundary of the exponential BCS regime (\ref{Tcas}).
At large values of $g$ (not shown) all curves $T_c(g)$
approach the asymptotic line $T_c =g/4$.
}
\label{TcFig}
\end{figure*}

Extrema $T_{*}$ of the function $g(T_c)$ (\ref{gTclarge})
are determined by the transcendental equation
\begin{equation}
\label{Tstar2trans}
e^{-g_{\lambda}/2T_{*}}= \frac{\lambda \pi}{3 g_{\lambda} \tilde J^2}
T_{*}^3
\end{equation}
This equation always has a trivial solution $T_{*}^{\prime}=0$
corresponding to the (local) maximum of $g(T_c)$. This is the QCP,
and the curve $T_c(g)$ approaches the QCP normally to the abscissa
(see Fig.\ \ref{TcFig}). Two non-trivial solutions of (\ref{Tstar2trans})
exist if the couplings satisfy the condition
\begin{equation}
\label{mincondI}
\tilde J > \mathbb{C}_1 \lambda^\frac12 g_{\lambda}~,
\end{equation}
where
\begin{equation}
\label{C1}
\mathbb{C}_1 \equiv  \sqrt{\frac{\pi}{24}\Big(\frac{e}{3}\Big)^3}
\approx 0.3121~.
\end{equation}
There is only one solution within the validity region of the
approximation (\ref{gTclarge}), and it corresponds to the
minimum of $g(T_c)$.
If the couplings satisfy (\ref{mincondI}) then
\begin{eqnarray}
\label{ua0}
u &\equiv& \ln 3 +\frac13 \ln a_{\circ} >1 \\ \nonumber
a_{\circ} &\equiv& \frac{24 \tilde J^2}{\pi \lambda g_{\lambda}^2}~,
\end{eqnarray}
and the minimum can be found analytically as
\begin{equation}
\label{Tstar2}
T_{*} \approx \frac{g_{\lambda}}{6}
\bigg( \frac{1}{u} -\frac{\ln u}{u^2} \bigg)~.
\end{equation}
For the left border of the re-entrant region we obtain
\begin{equation}
\label{gmin2}
g_{\textrm{min}} = g_{\lambda}-
\frac{\lambda \pi}{3  \tilde J^2}T_{*}^2+
\frac{4\lambda \pi}{3 g_{\lambda} \tilde J^2} T_{*}^3
\end{equation}
The above equations agree well with the numerical solutions of
the MFA (\ref{EqMF}) at different values of couplings (from
comparison of the asymptotics (\ref{gTcsmall}, \ref{gTclarge})
and the numerical curves at various couplings and temperatures
we found the deviations $\sim 5 \%$ at most). More importantly,
the analytical results of this subsection allows us to
understand in details the interplay of the scales provided by model's
couplings and the temperature, resulting in the re-entrance.
Let us explain this on the example of two characteristic
numerical curves shown in Fig.\ \ref{TcFig}.

The  curve shown for $J=0,\lambda =1,\varepsilon=0$ ($\tilde J=0.25$)
corresponds to the case of small $\tilde J$.
At $T_c \agt \tilde J/2=0.125$ it is well described by the equations
for the regime (\ref{tilbetsmall}). Its re-entrant behavior and,
in particular, the minimum $g_{\textrm{min}}$ is due to the last
term on the r.h.s. of (\ref{gTcsmall}).
At lower temperatures $T_c \alt 0.125$ the asymptotics
(\ref{gTcsmall}) is not applicable, the curve is described by
(\ref{gTclarge}). Note that since
$\mathbb{C}_1 \lambda^\frac12 g_{\lambda} \approx 0.8229$,
the condition (\ref{mincondI}) for the minimum is broken, and
the asymptotics (\ref{gTclarge}) describes the featureless
low-temperature evolution of this curve towards the maximum at the QCP.

The second curve in Fig.\ \ref{TcFig} with
$J=0.75, \lambda =1, \varepsilon=0$ ($\tilde J=1$) corresponds to
the case of large $\tilde J$. The whole re-entrant region ($T_c<0.4$),
including the position of the minimum $g_{\textrm{min}}$
($\mathbb{C}_1 \lambda^\frac12 g_{\lambda} \approx 0.8229$) is
described by the interplay of the last two terms on the
r.h.s. of Eq.(\ref{gTclarge}).
Comparison of Eqs.(\ref{gmin1},\ref{gmin2}) allows also to understand
a more pronounced re-entrant behavior for the case of smaller $J$.

As we see from our analysis of Eqs.(\ref{gTcsmall},\ref{gTclarge}) in
the both regimes (\ref{tilbetsmall},\ref{tilbetlarge}), the
the re-entrant behavior on the phase diagram occurs at
any $\lambda \neq 0$.
%
\subsubsection{Case $\varepsilon \neq 0$; double re-entrance,
no QCP}\label{epsneq0}
%
The absence of re-entrance at $\lambda =0$ can be proven rigorously.
Indeed, combining Eqs.(\ref{mzbe},\ref{mzTc}) we obtain
the equation
\begin{equation}
\label{mziplL0}
m_z^{-1}=g+ \frac{4 \varepsilon^2}{\pi J}
\eta_n \big( 0,\frac{J}{2} \ln \frac{1+2 m_z}{1-2 m_z} \big)
\end{equation}
which has one and only one solution $m_z \in [0,1/2]$ for a given
value of $g$. This solution in its turn provides a unique value of
$T_c$ via Eq.(\ref{mzTc}), thus no re-entrance.

At $\lambda \neq 0$ continuous evolution of $T_c(g)$ between the regimes
of strong Ising coupling and BCS- [cf. Eq.(\ref{Tcas})] can occur either
with a double re-entrance [i.e., with one minimum and one maximum
of $g(T_c)$] within the re-entrant region
\begin{equation}
\label{reentrII}
g_{\textrm{min}} <g<g_{\textrm{max}}~,
\end{equation}
or without re-entrance. In the latter case the
function $T_c(g)$ [or $g(T_c)$] has only an inflexion point
(see Fig.\ \ref{TcFig}).

Following the analysis given in the previous subsection,
we obtain for the case of
small $\tilde J$ in the regime (\ref{tilbetsmall})
\begin{equation}
\label{gTcsmalleps}
g = 2+4 e^{-1/T_c} +\frac{\lambda \tilde J-2 \varepsilon^2}{4 T_c}
\end{equation}
Again, the re-entrant behavior is conditioned by
the existence of a minimum of $g(T_c)$. It exists if, at least
\begin{equation}
\label{epsCondI}
\varepsilon < \frac{1}{\sqrt{2}} \sqrt{\lambda \tilde J}
\approx 0.7071 \sqrt{\lambda \tilde J}
\end{equation}
The unique minimum of $g(T_c)$ (\ref{gTcsmalleps})
defines the left border of the re-entrance region
$g_{\textrm{min}}$ and
corresponds to the critical temperature $T_{*}$ given by
Eqs.(\ref{Tstar1},\ref{gmin1}) with $\kappa_{\circ} \mapsto \kappa$,
and
\begin{equation}
\label{kappa}
\kappa \equiv \frac{16}{\lambda \tilde J-2 \varepsilon^2}
\end{equation}
The consistency imposes the constraint analogous to
(\ref{Condri}), more stringent than the ``minimal requirement"
(\ref{epsCondI}).

The other regime (\ref{tilbetlarge}) is described by the approximation
\begin{equation}
\label{gTclargeEps}
g =g_{\lambda}+4 e^{-g_{\lambda}/2T_c} -
\frac{\pi T_c^2}{3 \tilde J^3}
\big( \lambda \tilde J  + \varepsilon^2 \big)-
\frac{4 \varepsilon^2}{\pi \tilde J} \ln \frac{\mathbb A \tilde J}{2 T_c}
\end{equation}
As we have explained in the previous subsection
for the case $\varepsilon=0$, the asymptotics (\ref{gTcsmalleps})
is applicable only for small $\tilde J$ at the intermediate temperatures
(\ref{tilbetsmall}), while (\ref{gTclargeEps}) can be applied at arbitrary
low temperatures, including the BCS region. The latter, given by the
exponential dependence in (\ref{Tcas}), can be recovered if we retain
only the first and last (leading) terms on the r.h.s. of
Eq.(\ref{gTclargeEps}).

In the regime (\ref{tilbetlarge}) the re-entrance occurs if the equation
for extrema of (\ref{gTclargeEps})
\begin{equation}
\label{Trans2}
e^{-g_{\lambda}/2T_{*}}=
\frac{\pi T_{*} }{3 g_{\lambda} \tilde J^3}
\Big[\big( \lambda \tilde J  + \varepsilon^2 \big) T_{*}^2-
\frac{6 \varepsilon^2 \tilde J^2}{\pi^2} \Big]
\end{equation}
has non-trivial solutions. Note in making comparison of
Eq.(\ref{Trans2}) to its counterpart (\ref{Tstar2trans}) at
$\varepsilon=0$, that coupling $\varepsilon$ destroys the QCP
\cite{CGbig03}, as immediately seen from (\ref{gTclargeEps}). So the
trivial solution $T_{*}=0$ of Eq.(\ref{Trans2}) corresponds to the
unphysical singularity of $g$. As follows from
(\ref{Trans2},\ref{tilbetlarge}), non-trivial solutions are possible
if, at least
\begin{equation}
\label{epsCondII}
\varepsilon < \frac{\pi}{2\sqrt{6}}
\frac{1}{\sqrt{1-\frac{\pi^2}{24}}}  \sqrt{\lambda \tilde J}
\approx 0.8357 \sqrt{\lambda \tilde J}
\end{equation}
If this condition is satisfied, the transcendental
equation (\ref{Trans2}) has at least one solution, corresponding to
maximum of $g(T_c)$. To leading order in $\varepsilon$,
it occurs at the temperature
\begin{equation}
\label{Tstmax}
T_{*}^{\prime} \approx \frac{\sqrt{6}}{\pi}
\Big( \frac{\tilde J}{\lambda} \Big)^{\frac12}
\varepsilon
\end{equation}
and the coupling
\begin{equation}
\label{gmax}
g_{\textrm{max}} \approx  g_{\lambda}-
\frac{4 \varepsilon^2}{\pi \tilde J}
\ln \frac{\mathbb A \pi \sqrt{\lambda \tilde J}}
{2 \sqrt{6} \varepsilon}
\end{equation}
If the couplings meet both the conditions (\ref{mincondI}) and
\begin{equation}
\label{mincond}
\varepsilon < \mathbb{C}_2 g_{\lambda}
\Big( \frac{\lambda}{\tilde J} \Big)^\frac12 ~,
\end{equation}
where
\begin{equation}
\label{C2}
\mathbb{C}_2 \equiv \frac{\pi}{2} \sqrt{ \frac{47-13 \sqrt{13}}{36}}
\approx 0.0936 ~,
\end{equation}
then a second solution of (\ref{Trans2}) ($T_{*}^{\prime \prime}$),
corresponding to minimum of $g(T_c)$ exists. This minimum, located
between
\begin{equation}
\label{Tminloc}
T_{*}^{\prime } < T_{*}^{\prime \prime} < \frac{g_{\lambda}}{6}
\end{equation}
is given approximately by Eq.(\ref{Tstar2}) where $u$
in (\ref{ua0}) is modified by  $a_{\circ} \mapsto a$,
and
\begin{equation}
\label{a}
a \equiv
\frac{24 \tilde J^3}{\pi g_{\lambda}^2(\lambda \tilde J +\varepsilon^2)}
\end{equation}
For validity of expansion (\ref{Tstar2}) we assume $u>1$.

The analytical results of this subsection allows us to describe the
behavior of the coupled model at $\varepsilon \neq 0$, following
from the MFA equations (\ref{EqMF}), both qualitatively and
quantitatively. In Fig.\ \ref{TcFig} two counterparts ($\varepsilon=0.1$)
of the numerical curves discussed in the previous subsection are shown.
For this case of small $\varepsilon$, re-entrance is possible,
according to conditions (\ref{epsCondI},\ref{epsCondII}).
The re-entrant behavior at the temperatures $T_c \agt T_{*}^{\prime}$
is not modified essentially by the presence of new coupling
$\varepsilon$ comparatively to the case $\varepsilon=0$, and is in fact
controlled by couplings $J, \lambda, g$.  Since we have already discussed
it in detail for the case $\varepsilon=0$, we will not dwell on it any
more. In the low-temperature regime (\ref{gTclargeEps}) coupling
$\varepsilon$ changes drastically the behavior of $g(T_c)$ at
$T_c \alt T_{*}^{\prime}$, creating a maximum at $g(T_{*}^{\prime})$
described well by the approximation (\ref{Tstmax}) and turning
$g(T_c)$ away from the QCP towards the BCS region.  For the BCS
regime Eq.(\ref{Tcas}) provides a virtually exact solution.

As follows from the inequalities
(\ref{epsCondI},\ref{epsCondII},\ref{mincond}) an
increase of $\varepsilon$ can suppress the re-entrance even at
$\lambda \neq 0$. The necessary conditions
(\ref{epsCondI},\ref{epsCondII}) for extrema of the two asymptotics
(\ref{gTcsmalleps},\ref{gTclargeEps})
are close, albeit with a rather small mismatch of the coefficient.
Conditions for re-entrance are more stringent, since they
require the consistency between the solutions for
extrema and the validity ranges of the appropriate asymptotics.
The (overrated) critical value
$\varepsilon_{\circ} \sim 0.7 \sqrt{\lambda \tilde J}$ gives a good
simple estimate for the boundary where the re-entrance disappears from the
whole curve $g(T_c)$, whether $\tilde J >1$ or $\tilde J <1$.
An example of the curve $T_c(g)$ without re-entrance
is shown in Fig.\ \ref{TcFig}.

To summarize our analysis of the re-entrance for the cases
$\varepsilon=0$ and $\varepsilon \neq 0$: it reveals the robustness
of this phenomenon in the coupled model (\ref{Hamdls}) and its
underlying mechanism, namely, competition between different scales
defined by the  couplings $J, \lambda, \varepsilon, g$ and the
temperature. These competing scales (interactions) are not related
to the Ising frustration which is present in the model as well
[$(J_{\text{\tiny $\square$}},J_1,J_2)>0$], since the latter is not
accounted for explicitly by our mean-field equations. This competing
mechanism for the re-entrance appears to be robust and not being an
artifact of the MFA. Re-entrant phases due to competing interactions
are known also, e.g., from exact solution of the Ising model on the
union-jack lattice \cite{Vaks65}, or from analyses of decorated
Ising models \cite{Fradkin76}.

It is not clear for us at the moment how the proposed re-entrance
can be observed. $\rm NaV_2O_5$ is, up-to-date, the only known
compound with the spin-SAF transition and does not show re-entrance.
This is in agreement with our estimates for the parameters for the
effective Hamiltonian for this compound. They give its $g$ located
on the disordered side of the (destroyed) QCP, and the re-entrance
on the whole curve $T_c(g)$ would be only very weak (i.e., localized
near $g \sim g_{\lambda}$), if any. It appears experimentally that,
e.g., external pressure cannot modify the in-plane parameters of
$\rm NaV_2O_5$ strongly enough, such that re-entrance would be
generated. The variations of the interlayer couplings under
pressure, on the other hand, generate various types of order
(including devil's staircase) with regard to the plane stacking,
while the in-plane SAF order remains unaffected
\cite{Ohwada01,Stack04}.

%
%
\section{Summary and Discussion}\label{Concl}
%
We study the 2D Ising model on a square lattice with
nearest-neighbor ($J_{\text{\tiny $\square$}}$) and non-equal
next-nearest neighbor ($J_{1,2}$) interactions. The cases of
classical and quantum  models are considered.

We find the ground state phase diagram of the classical Ising model
at arbitrary $J_{\text{\tiny $\square$}}, J_1,J_2$. Along with the
three ordered phases --ferromagnetic (FM), antiferromagnetic (AF),
and SAF-- known for $J_1=J_2$ \cite{FanWu69}, in a more general case
$J_1 \neq J_2$ there is a region of the coupling space with the
super-ferro-antiferromagnetic (SFAF) [or $(4\times 4)$] ground state
phase and an incommensurate (IC) phase at finite temperature, not
reported before.  The three phases -- SAF, SFAF, IC -- can occur
only in the presence of competing interactions (frustrations) on the
Ising model's plaquette.

\begin{figure*}
\includegraphics[width=10.0cm]{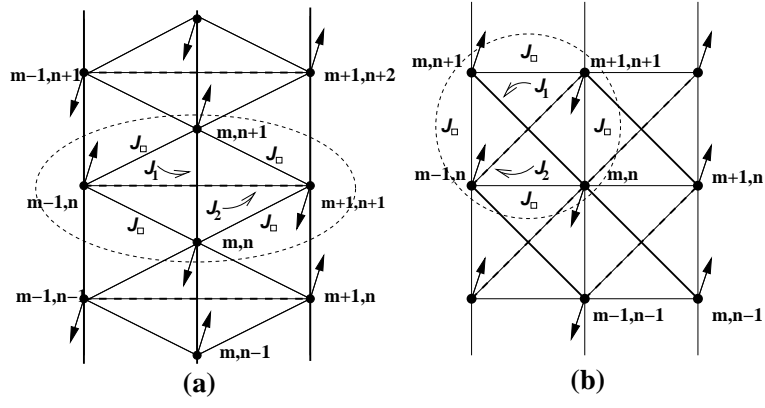}
\caption{ (a): 2D effective lattice of coupled ladders. A vertical
line and a dot represent a single ladder and its rung where
pseudospin $\bm{\mathcal T}_{mn}$ and spin $\mathbf S_{mn}$ (not
shown) reside. In the region encircled by the dashed line the Ising
couplings between pseudospins are indicated. Two pseudospins from
$(m-1)$-th and $(m+1)$-th ladders are coupled not only by $J_2$
(bold dashed line), but also via the dimerization constant
$\varepsilon$. The ${\mathcal T}_{mn}^x$-ordering pattern shown
corresponds to the SAF phase. (b): The same after mapping on a
square lattice. } \label{LatMapFig}
\end{figure*}

A particularly interesting conclusion from the analysis of the
quantum model's phase boundaries is that transverse field ($\Omega$)
can stabilize the IC ground-state phase (located for $\Omega =0$ in
the region with $\textrm{sign}(J_1/J_2)=-1$) in some parts of the AF
and SAF regions of the coupling space where $(J_1,J_2)>0$, but $J_1
\neq J_2$. These regions, along with vicinities of the special
planes of degeneracy (triangulation and frustration) in coupling
space, are good candidates for the quantum model to demonstrate a
very non-trivial critical behavior. Leaving this for a future work,
we hope that our findings will inspire additional interest in this
model. Taking into account only one simple example of a mapping
shown in Fig.\ \ref{LatMapFig}, it is clear the model with $J_1 \neq
J_2$ is not so exotic.

We analyze the IMTF coupled to the $XY$ spin chains in the
restricted (SAF) region of $(J_{\text{\tiny $\square$}}, J_1,J_2)$
where the IMTF has a simple two-phase (disordered-SAF) diagram with
a QCP, similar to that of the transverse nn model. Our interest in
this model is motivated by the problem of the phase transition in
the quarter-filled ladder compound $\rm NaV_2O_5$. The predictions
of the mean-field equations for the critical properties of the
coupled spin-pseudospin model do not differ essentially from our
earlier results for a simpler Hamiltonian \cite{CGbig03}. Due to the
spin-pseudospin coupling $\varepsilon$, the QCP of the nn and nnn
IMTF is destroyed, and in the whole SAF region (\ref{SAFsafe}) of
Ising couplings the spin-pseudospin model undergoes the spin-SAF
transition. We should point out, that albeit the exponential BCS
regime (\ref{Tcas}) on the disordered side of the IMTF QCP
$g<g_{\lambda}$ formally extends up to $g=0$, decreasing $g$, i.e.
$J_1+J_2$, will eventually remove us from the coupling region
(\ref{SAFsafe}) where the SAF pseudospin solution of the MFA
equations is applicable.

We  perform a detailed analytical study of re-entrance in the
coupled model. In particular, we establish the conditions when it
can occur. The analytical results not only agree well with the
direct numerical calculations in various regimes, but allows us to
understand the physical mechanism of re-entrance due to interplay of
competing interactions in the coupled model.

In this work we gain more insights on the transition in the
spin-pseudospin model, and we can sharpen our previous statements
concerning the applications to $\rm NaV_2O_5$ \cite{CGbig03}. The
present analysis of Ising sector allows us to identify the 2D
long-range charge order in that compound as the SAF phase. As
follows from known results on the ordering of the frustrated 2D
Ising model into the SAF phase \cite{Liebmann86}, the spin-SAF
transition has non-universal coupling-dependent critical indices.
Experiments indicate rather wide regions of the two-dimensional
structural (charge-ordering) fluctuations characterized by the
critical index  $\beta \approx 0.17-0.19$
\cite{Ravy99,Gaulin00,Fagot00}, close to $\beta = 1/8$ of the 2D
Ising model. Due to known difficulties in extracting critical
indices from experimental data, it seems problematic to diagnose the
deviations from universality caused by $0< J_{\text{\tiny
$\square$}} /J_1<1$. (Note that in the limit $J_{\text{\tiny
$\square$}} \to 0$ the PM-SAF transition enters into the 2D Ising
universality class, and for $\rm NaV_2O_5$, due to its geometry, the
ratio $J_{\text{\tiny $\square$}}/J_1$ should be small.)

On the theory side, the critical indices of the PM-SAF transition as
functions of couplings in the nn and nnn Ising model have been
calculated by various methods only at $J_1 = J_2$ \cite{Liebmann86}.
The critical exponents are unknown for the case $J_1 \neq J_2$, and
it appears to be an interesting problem to study.

Another very interesting issue we addressed recently in a separate
study \cite{Stack04}, is the 3D nature of the transition in $\rm
NaV_2O_5$. According to the correlation lengths measurements
\cite{Ravy99}, upon approaching $T_c =34K$ the $2D \to 3D$ crossover
of the pretransitional structural fluctuations occurs somewhere at
$T \sim 50K$. The model considered in the present work deals with a
single plane, leaving aside the question of charge ordering along
the third (stacking) direction. The phase transition in $\rm
NaV_2O_5$ quadruples the unit cell in the stacking direction, and
the recent X-ray experiments, carried out deep in the ordered phase
\cite{Sma02,Grenier02} revealed peculiar stacking ordering patterns
of the super-antiferroelectrically charge-ordered planes. In
addition, the pressure can change these patterns and even generate a
multitude of higher-order commensurate superstructures in the
stacking direction (devil's staircase) \cite{Ohwada01}. To explain
these phenomena we proposed a 3D extension of the Ising sector with
additional competing couplings between the nearest and next-nearest
planes \cite{Stack04}. In the limit $J_{\text{\tiny $\square$}} \to
0$ the Ising sector reduces to two identical interpenetrating
decoupled 3D ANNNI models. Although inclusion of the competing
interlayer couplings accommodates the explanation for the observed
stacking charge order in the framework of the spin-SAF (in-plane)
mechanism of the transition in $\rm NaV_2O_5$, a deeper
understanding of the critical properties of a very complicated model
with the 3D Ising sector warrants a further work.
%
%
%
\begin{acknowledgments}
We are grateful to D.I. Khomskii, E. Orignac, and P.N. Timonin for
helpful discussions. This work is supported by the German Science
Foundation. G.Y.C. also acknowledges support from a LURF grant.
\end{acknowledgments}
%


\begin{thebibliography}{}
%
\bibitem{Liebmann86} R. Liebmann,
\textit{Statistical Mechanics of Periodic Frustrated Ising Systems}
(Springer, Berlin, 1986).
%
\bibitem{SachdevQPT} S. Sachdev,
\textit{Quantum Phase Transitions} (Cambridge University Press,
Cambridge, 1999).
%
\bibitem{Chak96} B.K. Chakrabarti, A. Dutta, and P. Sen,
\textit{Quantum Ising Phases and Transitions in Transverse Ising Models}
(Springer, Berlin, 1996).
%
\bibitem{Moessner01} R. Moessner and S.L. Sondhi,
Phys. Rev. B \textbf{63}, 224401 (2001).
%
\bibitem{Most03} M.V. Mostovoy, D.I. Khomskii, J.Knoester, and
N.V. Prokof'ev, Phys. Rev. Lett. {\bf 90}, 147203 (2003).
%
\bibitem{CGbig03} G.Y. Chitov and C. Gros,
Phys. Rev. B \textbf{69}, 104423 (2004).
%
\bibitem{Lem03} P. Lemmens, G. G\"untherodt, and C. Gros,
Phys. Reports. {\bf 375}, 1 (2003).
%
\bibitem{FanWu69} C. Fan and F.Y. Wu,
Phys. Rev. {\bf 179}, 560 (1969).
%
\bibitem{Landau85} Such state, with the name of $(4 \times 4)$
superstructure was described before, e.g., by D.P. Landau and  K.
Binder, Phys. Rev. B \textbf{31}, 5946 (1985). It can also occur in
the 2D Ising model where along with nn and nnn, the third-neighbor
interactions are included.
%
\bibitem{Bak82} P. Bak, Rep. Prog. Phys. {\bf 45}, 587 (1982).
%
\bibitem{Selke88} W. Selke, Phys. Rep.  {\bf 170}, 213 (1988).
%
\bibitem{KriMu77} S. Krinsky and D. Mukamel,
Phys. Rev. B \textbf{16}, 2313 (1977).
%
\bibitem{Dom78} E. Domany, M. Schick, J.S. Walker, and
R.B. Griffiths, Phys. Rev. B \textbf{18}, 2209 (1978).
%
\bibitem{Wannier50} G.H. Wannier, Phys. Rev. {\bf 79}, 357 (1950);
Errata: Phys. Rev. B \textbf{7}, 5017 (1973).
%
\bibitem{Houtappel50} R.M.F. Houtappel, Physica {\bf 16}, 425 (1950).
%
\bibitem{StephensonIII} J. Stephenson,
J. Math. Phys. {\bf 11}, 413 (1970).
%
\bibitem{StephensonIV} J. Stephenson,
J. Math. Phys. {\bf 11}, 420 (1970).
%
\bibitem{Copper82/1} S.N. Coppersmith, D.S. Fisher, B.I. Halperin,
P.A. Lee, and W.F. Brinkman, Phys. Rev. B \textbf{25}, 349 (1982);
Phys. Rev. Lett. {\bf 46}, 549 (1981); Erratum:
Phys. Rev. Lett. {\bf 46}, 869 (1981).
%
\bibitem{Kugel82} K.I. Kugel and D.I. Khomskii,
Usp. Fiz. Nauk \textbf{136}, 621 (1982);
[Sov. Phys. Usp. \textbf{25}(4), 231 (1982)].
%
\bibitem{Most98} M.V. Mostovoy and D.I. Khomskii, Solid St. Comm.
\textbf{113}, 159 (1999).
%
\bibitem{Sa00} D. Sa and C. Gros, Eur. Phys. J. B \textbf{18}, 421 (2000).
%
\bibitem{Most02} M.V. Mostovoy, D.I. Khomskii, and J. Knoester,
Phys. Rev. B \textbf{65}, 064412 (2002).
%
\bibitem{Grenier02} S. Grenier, A. Toader, J. E. Lorenzo, Y. Joly,
B. Grenier, S. Ravy, L. P. Regnault, H. Renevier, J. Y. Henry, J.
Jegoudez, and A. Revcolevschi, Phys. Rev. B \textbf{65}, 180101(R)
(2002).
%
\bibitem{Stack04} G.Y. Chitov and C. Gros, J. Phys.: Condens. Matter
\textbf{16}, L415 (2004).
%
\bibitem{Diag05} C. Gros and G.Y. Chitov,
Europhys.  Lett. \textbf{69}, 447 (2005).
%
\bibitem{Blinc74} R. Blinc and B. \v{Z}ek\v{s},
\textit{Soft Modes in Ferroelectrics and Antiferroelectrics},
(North-Holland Publishing Co., Amsterdam, 1974).
%
\bibitem{Vaks65} V.G. Vaks, A.I. Larkin, and Y.N. Ovchinnikov,
Zh. Eksp. Teor. Fiz. \textbf{49}, 1180 (1965)
[Sov. Phys. JETP, \textbf{22}, 820 (1966)].
%
\bibitem{Fradkin76} E.H. Fradkin and T.P. Eggarter,
Phys. Rev. A \textbf{14}, 495 (1976).
%
\bibitem{Ohwada01}
K. Ohwada, Y. Fujii, N. Takesue, M. Isobe, Y. Ueda, H. Nakao,
Y. Wakabayashi, Y. Murakami,  K. Ito, Y. Amemiya,  H. Fujihisa,
K. Aoki, T. Shobu, Y. Noda, and N. Ikeda,
Phys. Rev. Lett. \textbf{87}, 086402 (2001).
%
\bibitem{Ravy99} S. Ravy, J. Jegoudez, and A. Revcolevschi,
Phys. Rev. B \textbf{59}, 681 (1999).
%
\bibitem{Gaulin00} B.D. Gaulin, M.D. Lumsden, R.K. Kremer,
M.A. Lumsden, and H. Dabkowska,
Phys. Rev. Lett. \textbf{84}, 3446 (2000).
%
\bibitem{Fagot00} Y. Fagot-Revurat, M. Mehring, and R.K. Kremer,
Phys. Rev. Lett. \textbf{84}, 4176 (2000).
%
\bibitem{Sma02} S. van Smaalen, P. Daniels, L. Palatinus,
and R.K. Kremer, Phys. Rev. B \textbf{65}, 060101 (2002).
%
\end{thebibliography}
\end{document}